\documentclass[fp,twocolumn]{jpsj3}
\usepackage{txfonts}
\usepackage{url}
\usepackage{bm}
\usepackage{url}
\usepackage{color}
\usepackage[dvipdfmx]{graphicx}
\usepackage{here}
\usepackage{widetext}

\title{Perfect transmission and perfect reflection of Bogoliubov quasiparticles in a dynamically unstable Bose-Einstein condensate}
\author{Terumichi Ohashi and Yuki Kawaguchi\thanks{kawaguchi@nuap.nagoya-u.ac.jp}}
\inst{Department of Applied Physics, Nagoya University, Nagoya 464-8603, Japan} 

\abst{
The Nambu-Goldstone (NG) mode in a Bose-Einstein condensate (BEC) transmits a potential barrier with probability 1 in the zero-energy limit, which is known as the anomalous tunneling.
In this paper, we investigate the tunneling properties of quasiparticles in a dynamically unstable BEC. 
We prepare a multi-component BEC (binary and spin-1 BEC) in a dynamically unstable state and solve the tunneling problem of the spin-wave excitation from the condensate.
We find that the perfect transmission occurs even when the BEC is dynamically unstable if the spin-wave is the NG mode.
Here, the mode that exhibits the perfect transmission is the dynamically unstable spin-wave mode, which is a pure-imaginary-eigenvalue solution of the Bogoliubov-de Gennes equation.
Hence, we should take the zero-energy limit along not the real axis but the imaginary axis.
We also demonstrate the existence of the perfect reflection of a dynamically unstable mode at the point where the imaginary part of the eigenvalue takes its maximum. In this case, the incident and reflected waves destructively interfere, and the amplitude of the quasiparticle wave function is strongly suppressed.
We numerically confirm that the perfect reflection is a generic nature of dynamically unstable modes and not related to the NG mode.
}

\begin{document}
\maketitle
\section{Introduction}
The concept of elementary excitations plays a key role in understanding fundamental properties of quantum many-body systems, ranging from ground-state property to non-equilibrium transport phenomena~\cite{P.W.Anderson}. 
In particular, in a system with spontaneous symmetry breaking, the nature of the system at low energy is dominated by the gapless Nambu-Goldstone (NG) mode associated with the broken symmetry.
Examples include phonons in a Bose-Einstein condensate (BEC) and magnons in a ferromagnet,
which are associated with the breaking of the U(1) gauge symmetry and the SO(3) spin-rotational symmetry, respectively.

\begin{figure}
	\centering
    \includegraphics[clip,width=8.5cm]{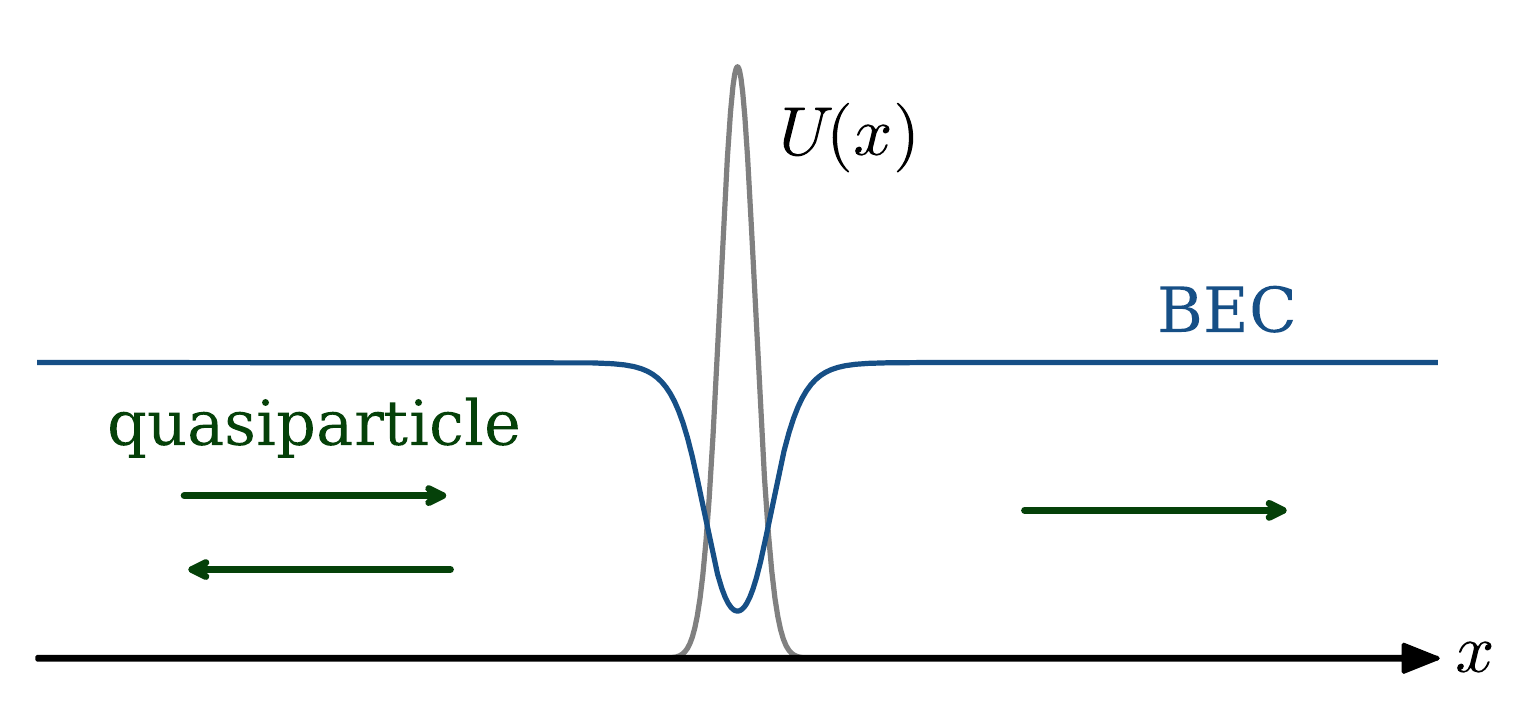}	
		\caption{
		Schematic of the tunneling problem of a Bogoliubov quasiparticle in a BEC under a potential barrier $U(\bm x)$.
        A quasiparticle with energy $E$ injected from the left is reflected to the left or transmitted to the right.
        Whereas a non-interacting particle is reflected with probability 1 in the low energy limit $E\to 0$,
        the Bogoliubov quasiparticle associated with the spontaneous symmetry breaking exhibits the perfect transmission at $E\to 0$.
        This perfect transmission is called the anomalous tunneling~\cite{KOVRIZHIN2001392,PhysRevLett.90.130402}.
        }
	\label{fig:schematic_potential}
\end{figure}

Anomalous tunneling is one of the salient features of the elementary excitations in a BEC, which is the phenomenon that a quasiparticle in a BEC transmits a potential barrier with probability 1 in the low energy limit~\cite{KOVRIZHIN2001392,PhysRevLett.90.130402} (see Fig.~\ref{fig:schematic_potential}). 
This behavior is the opposite of the tunneling problem of non-interacting particles described by the Schr\"odinger equation, where the transmission probability goes to zero in the low energy limit.
So far, the anomalous tunneling has been studied in
BECs with supercurrent flow~\cite{Danshita_2006,PhysRevA.78.043601,doi:10.1143/JPSJ.78.023001}, at finite temperature~\cite{doi:10.1143/JPSJ.77.013602}, and with spin degrees of freedom~\cite{PhysRevA.83.033627,PhysRevA.83.053624,PhysRevA.84.013616}.
The relations with Josephson current~\cite{PhysRevA.78.013628,PhysRevA.79.063619},
impedance matching~\cite{PhysRevA.78.063611},
and scattering at magnetic domain wall and at impurities ~\cite{PhysRevA.86.023622,Watabe_2015}
were also discussed.
The above works revealed that
the anomalous tunneling occurs in the zero-energy limit of the NG mode: The NG boson in the zero-energy limit is identical to the condensed particles, and hence, its wave function extends over the both side of the barrier potential, resulting in the perfect transmission.
Apart from the zero-energy limit of the NG mode, the Fano resonance between the NG and Higgs modes and the perfect transmission of the Higgs mode via antibound states have been recently  predicted~\cite{PhysRevA.92.043610,PhysRevA.100.063612}.

This paper is motivated by the result in Refs.~\citen{Danshita_2006,PhysRevA.78.043601,doi:10.1143/JPSJ.78.023001} that the perfect transmission does not occur in a scalar BEC flowing with the critical current at the onset of the Landau instability. It is well known that there are two types of instabilities in a BEC; the Landau instability and the dynamical instability. The former is the energetical instability characterized by a negative excitation energy, 
whereas the latter is the instability against the exponential growth of zero-energy excitations that are characterized by complex eigenfrequencies~\cite{PhysRevLett.93.140406,PhysRevLett.93.160406,sadler2006spontaneous,KAWAGUCHI2012253}.
It is then natural to ask how the anomalous tunneling occurs in a dynamically unstable system, which is yet to be investigated.

%

In this paper, we investigate the tunneling properties of quasiparticles in the presence of dynamically unstable modes. 
When a prepared BEC is dynamically unstable, the Bogoliubov-de Gennes (BdG) equation that describes the excitation spectrum from the condensate has complex eigenvalues.
The tunneling problem is well-defined even in such a case: We solve the BdG equation with a given eigenvalue $E$ (which can be a complex value) and divide the obtained wave function into the incident, reflected, and transmitted waves, obtaining the reflection and transmission probabilities.

Below, we consider two situations: a completely-mixed binary (pseudo-spin-1/2) BEC and a spin-1 polar BEC.
Dynamical instabilities in such systems are experimentally observed in Refs.~\citen{PhysRevLett.82.2228,sadler2006spontaneous,PhysRevLett.115.245301,PhysRevLett.119.185302}.
Here, we choose the spin configuration of the condensate such that the system can be stable or dynamically unstable depending on the spin-dependent interaction parameter.
The BdG equations for spin waves in these systems are the same except for the quadratic Zeeman energy ($q_z$) term that appears only in the case of the polar BEC.
Since the $q_z$ term breaks the spin rotational symmetry, the spin-wave mode in a polar BEC is not the NG mode whereas it in a binary BEC is the NG mode.
We find that the perfect transmission occurs in the case of a binary BEC even when the system is dynamically unstable.
However, the zero-energy limit should be taken along not the real $E$ axis but the imaginary $E$ axis
so that the quasiparticle wave function coincides with the condensate one in the $E\to 0$ limit.
On the other hand, the perfect transmission does not occur for any parameters in a spin-1 polar BEC with $q_z\neq 0$
because the spin-wave mode is not the NG mode.
Instead, when the kinetic energy of the incident wave matches with the energy in the long-wavelength limit, the transmission probability resonantly increases.
For the case of a spin-1 polar BEC, the eigenvalue $E$ remains nonzero in the long-wavelength limit due to the quadratic Zeeman energy. We find that the transmission probability resonantly increases when the energy of the incident quasiparticle matches with this energy.

We also find that the perfect reflection occurs when ${\rm Im}E\neq 0$ and $d{\rm Im}E/dk=0$, where $k$ is the momentum of the incident wave.
This is understood as the consequence of the disappearance of the linearly independent plane-wave solution for the given $E$.
The fact that the perfect reflection occurs in both systems of a binary BEC and a spin-1 polar BEC
indicates that the origin of the perfect reflection is not related to the NG mode.

The organization of this paper is as follows.
In Sec.~\ref{sec:binary_BEC}, we investigate the tunneling properties of the spin-wave mode in a binary BEC: We introduce the system in Sec.~\ref{sec:model_of_a_binayBEC}, and derive solutions of the BdG equation in the absence of the barrier potential in Sec.~\ref{sec:Bogoliubov_spectrum_binayBEC}; In Sec.~\ref{sec:tunneling properties_magnon}, we solve the BdG equation under a barrier potential and discuss the tunneling properties.
In Sec.~\ref{sec:spin1_BEC}, we discuss the tunneling problem in a spin-1 polar BEC in the same manner as Sec.~\ref{sec:binary_BEC}.
Section~\ref{sec:conclusion} concludes the paper.

\section{Binary BEC \label{sec:binary_BEC}}
\subsection{Model\label{sec:model_of_a_binayBEC}}
We consider a binary BEC under a barrier potential at $x=0$ (Fig.~\ref{fig:schematic_potential}) and examine the tunneling properties of the quasiparticles through the barrier.
The energy functional of the system is given by
\begin{align}
    \mathcal{E}_{\rm binary}=&\int d\bm x\sum_{m}\left[\frac{\hbar^2}{2M}\Big|\frac{\partial}{\partial x}\Psi_m\Big|^2+U(x)|\Psi_m|^2\right] \nonumber\\
     &+\int d\bm x \Big[\sum_{m}\frac{g}{2}|\Psi_m|^4+ g'|\Psi_1|^2|\Psi_2|^2\Big],
	\label{eq:Energy_fuc_binaryBEC}
\end{align}
where $\Psi_m (m=1,2)$ is the condensate wave function, $M$ is the atomic mass, $U(x)$ is the barrier potential, and $g>0$ and $g'>0$ are the intra- and inter-species interaction strengths, respectively. For the sake of simplicity, 
we assume that the condensate wave function and the barrier potential depend only on $x$ in a three-dimensional system
so that the problem becomes essentially one dimensional.
We also assume 
that the strengths of the intra-species interaction for the $m=1$ and $2$ components are the same. The two components are miscible (immiscible) for $g>g'$ ($g<g'$)~\cite{C.J.Pethick_and_H.Smith,L.Pitaevskii_and_S.Stringari}.
The potential $U(x)$ goes to zero and the condensate density $n(x)=\sum_m|\Psi_m|^2$ converges to a constant value $n_0$ at $x\to\pm\infty$.

Taking the functional derivative of the energy functional~\eqref{eq:Energy_fuc_binaryBEC} with respect to $\Psi^*_m$, we obtain the time-dependent Gross-Pitaevskii (GP) equation:
\begin{subequations}
\begin{align}
    i\hbar \frac{\partial\Psi_1}{\partial t} 
    =&\left[-\frac{\hbar^2}{2M}\frac{\partial^2}{\partial x^2}+U(x) 
    +g|\Psi_1|^2+g'|\Psi_2|^2\right]\Psi_1,\\
    i\hbar \frac{\partial\Psi_2}{\partial t} 
    =&\left[-\frac{\hbar^2}{2M}\frac{\partial^2}{\partial x^2}+U(x) 
    +g|\Psi_2|^2+g'|\Psi_1|^2\right]\Psi_2.
\end{align}
\label{eq:GP_binaryBEC}
\end{subequations}
Below, we consider an equal-population mixture of the two components and prepare a completely overlapped state: $\Psi_1(x)=\Psi_2(x)\equiv \Phi(x)/\sqrt{2}$,
where $\Phi(x)$ satisfies
\begin{align}
    \mu \Phi=\left[-\frac{\hbar^2}{2M}\frac{d^2}{dx^2}+U(x)+\frac{g+g'}{2}|\Phi|^2\right]\Phi.
    \label{eq:GP_binaryBEC_stationary}
\end{align}
with $\mu$ being the chemical potential.
The initial state is a stationary solution of Eq.~\eqref{eq:GP_binaryBEC} regardless of whether the binary mixture is miscible or immiscible. However, its stability drastically changes at $g=g'$: When $g>g'$ (miscible), the initial state is the ground state of the system, and hence, the quasiparticle eigenfrequencies are all positive real numbers;
When $g<g'$ (immiscible), the initial state is dynamically unstable against phase separation, and complex eigenfrequencies appear.

Since $U(x)\to 0$ and $|\Phi|^2\to n_0(=\textrm{const.})$ at $x\to\pm\infty$,
Eq.~\eqref{eq:GP_binaryBEC_stationary} leads to $\mu=(g+g')n_0/2$.
We describe the equations with rescaling the energy, length, and time scales by $\mu$, $\hbar/\sqrt{M\mu}$, and $\mu/\hbar$, respectively.
Introducing the interaction parameter $\beta$ ($-1\le\beta\le 1$) as
\begin{align}
g=(g+g')\frac{1+\beta}{2},\ \ 
g'=(g+g')\frac{1-\beta}{2}
\label{eq:def_alpha}
\end{align}
and rewriting the wave functions as $\Psi_{1,2}(x)=\sqrt{n_0}\psi_{1,2}(x)$ and $\Phi(x)=\sqrt{n_0}\phi(x)$,
the dimensionless forms of Eqs.~\eqref{eq:GP_binaryBEC} and \eqref{eq:GP_binaryBEC_stationary} are given by
\begin{subequations}
\begin{align}
    i\frac{\partial\psi_1}{\partial t} 
    =&\left[\mathcal{L}_0 
    +(1+\beta)|\psi_1|^2+(1-\beta)|\psi_2|^2\right]\psi_1,\\
    i\frac{\partial\psi_2}{\partial t} 
    =&\left[\mathcal{L}_0
    +(1+\beta)|\psi_2|^2+(1-\beta)|\psi_1|^2\right]\psi_2,
\end{align}
\label{eq:GP_binaryBEC2}
\end{subequations}
and
\begin{align}
    &\left(\mathcal{L}_0-1+|\phi|^2\right)\phi=0,
    \label{eq:GP_binaryBEC_stationary2}
\end{align}
where
\begin{align}
    \mathcal{L}_0=-\frac{1}{2}\frac{d^2}{dx^2}+U(x).
    \label{eq:def_L0}
\end{align}

The dynamics of quasiparticle excitations from a condensate at low temperature is well described by the Bogoliubov theory. 
The BdG equation for a binary BEC is obtained by substituting
\begin{align}
    \begin{pmatrix}\psi_1 \\ \psi_2 \end{pmatrix} 
    &= e^{-it} \left[  \frac{\psi_0}{\sqrt{2}}\begin{pmatrix} 1\\1 \end{pmatrix}
    + \bm u e^{-iE t}+\bm v^*e^{+iE^* t}\right]
	\label{eq:bdg_linearizing_binaryBEC}
\end{align}
into Eq.~\eqref{eq:GP_binaryBEC2} and linearizing the equation with respect to $\bm u$ and $\bm v$,
where $\bm u=(u_1(x),u_2(x))^{\rm T}$ and $\bm v=(v_1(x),v_2(x))^{\rm T}$ are two-component spinors. The resulting eigenvalue equation is given by
\begin{align}
  \begin{pmatrix}
     (\mathcal{L}_0-1){\bm 1} + H_1      & H_2             \\
    -H^*_2          & -[(\mathcal{L}_0-1){\bm 1} + H^*_1] 
  \end{pmatrix}
  \begin{pmatrix}
    \bm{u} \\ \bm{v}
  \end{pmatrix}
  =E
  \begin{pmatrix}
    \bm{u} \\ \bm{v}
  \end{pmatrix},
\label{eq:bdg_binary}
\end{align}
where ${\bm 1}$ is the $2\times 2$ identity matrix, and
\begin{align}
    H_1 &= \frac{|\phi|^2}{2}
    \begin{pmatrix}
    3+\beta & 1-\beta         \\
    1-\beta & 3+\beta 
    \end{pmatrix}
    , \\
    H_2 &=\frac{\phi^2}{2}
    \begin{pmatrix}
    1+\beta  & 1-\beta\\
    1-\beta & 1+\beta 
    \end{pmatrix}.
\end{align}

When the two components are completely overlapped, the density-wave and spin-wave modes are decoupled in the BdG equation.
Indeed, defining the density-wave (phonon) mode $u^{\rm d}=u_1+u_2, v^{\rm d}=v_1+v_2$ and the spin-wave (magnon) mode $u^{\rm s}=u_1-u_2, v^{\rm s}=v_1-v_2$, the BdG Eq.~\eqref{eq:bdg_binary} is divided into the following two equations:
\begin{widetext}
\begin{align}
    \begin{pmatrix}
    \mathcal{L}_0 -1 + 2|\phi|^2 & \phi^2 \\
    -(\phi^*)^2 & -(\mathcal{L}_0-1 +2|\phi|^2)
    \end{pmatrix}
    \begin{pmatrix} u^{\rm d} \\ v^{\rm d} \end{pmatrix}
    &=E^{\rm d}\begin{pmatrix} u^{\rm d} \\ v^{\rm d} \end{pmatrix},
    \label{eq:BdG_binary_phonon_x}\\
    \begin{pmatrix}
    \mathcal{L}_0 -1 + (1+\beta)|\phi|^2 & \beta\phi^2 \\
    -\beta(\phi^*)^2 & -[\mathcal{L}_0 -1 + (1+\beta)|\phi|^2]
    \end{pmatrix}
    \begin{pmatrix} u^{\rm s} \\ v^{\rm s} \end{pmatrix}
    &=E^{\rm s}\begin{pmatrix} u^{\rm s} \\ v^{\rm s} \end{pmatrix}.
    \label{eq:BdG_binary_magnon_x}
\end{align}
\end{widetext}
When $U(x)=0$, 
the spectra of these equations exhibit gapless linear dispersions, 
corresponding to the NG phonon and NG magnon modes associated with the breaking of the U(1) gauge symmetry and the SO(2) spin rotation symmetry, respectively. (The spin here means the pseudo-spin-1/2 of the binary system.) 
For the case of $U(x)\neq 0$, the existence of the NG modes is confirmed by the fact that 
$u^{\rm d,s}=-(v^{\rm d,s})^*=\phi$ are the eigen solutions of Eqs.~\eqref{eq:BdG_binary_phonon_x} and \eqref{eq:BdG_binary_magnon_x} with $E^{\rm d,s}=0$.

We note that the GP Eq.~\eqref{eq:GP_binaryBEC_stationary2} and the BdG Eq.~\eqref{eq:BdG_binary_phonon_x} for the phonon mode are identical to those for a scalar BEC. Hence, the tunneling problem is the same as the case of a scalar BEC discussed in Ref.~\citen{PhysRevLett.90.130402}.
We therefore discuss the tunneling property of the spin-wave mode in the rest of this section.

\subsection{Bogoliubov spectrum in a uniform system\label{sec:Bogoliubov_spectrum_binayBEC}}
Before solving the tunneling problem, we analytically derive asymptotic forms of the quasiparticle wave function at $x\to \pm\infty$.
For the case of $U(x)=0$, the stationary solution of Eq.~\eqref{eq:GP_binaryBEC_stationary2} is $\phi=1$.
Substituting $(u^{\rm s},v^{\rm s})^{\rm T}=(ue^{ikx},ve^{ikx})^{\rm T}$ and $\phi=1$ to Eq.~\eqref{eq:BdG_binary_magnon_x}, the BdG equation for spin-wave modes at $U(x)=0$ is given by
\begin{align}
    \begin{pmatrix}
    \epsilon_k+ \beta & \beta \\
    -\beta & -(\epsilon_k + \beta)
    \end{pmatrix}
    \begin{pmatrix} u \\ v \end{pmatrix}
    &=E^{\rm s}\begin{pmatrix} u \\ v \end{pmatrix},
    \label{eq:BdG_binary_magnon_k}
\end{align}
which has the eigenvalue
\begin{align}
    E^{\rm s}=\sqrt{\epsilon_k(\epsilon_k+2\beta)},
\label{eq:Bdg_binary_magnon_eigenvalue}
\end{align}
where $\epsilon_k\equiv k^2/2$.
Note that when $\beta<0$, $E^{\rm s}$ becomes pure imaginary for small $k$ and the system becomes dynamically unstable.
This instability is because the immiscible components [$g<g'$, which is equivalent to $\beta<0$, see Eq.~\eqref{eq:def_alpha}] are mixed in the initial state.
Our interest in this paper is how such dynamically unstable modes are reflected or transmitted by the barrier potential.
For the case of $\beta\ge 0$, for which the system is dynamically stable, almost the same situation has been discussed in the previous works~\cite{PhysRevA.83.053624,PhysRevA.84.013616}, where the tunneling problem of spin waves in a spin-1 polar state has been investigated and the perfect transmission was observed in the low energy limit.
The BdG Eq.~\eqref{eq:BdG_binary_phonon_x} for density-wave modes at $U(x)=0$
has the eigenvalue $E^{\rm d}=\sqrt{\epsilon_k(\epsilon_k+2)}$, which is positive real for $^\forall k$.

In the tunneling problem, we will solve the quasiparticle wave function injected from $x=-\infty$ with an energy $E$.
Here, we therefore calculate normalized eigenvectors of Eq.~\eqref{eq:BdG_binary_magnon_k} as a function of an incident eigenvalue $E$.
We should be careful to calculate the normalization constant for the BdG equation for a bosonic system since the BdG equation is a non-Hermitian matrix equation.
For the case of Eq.~\eqref{eq:BdG_binary_magnon_k}, the eigenvector is normalized as $|u|^2-|v|^2=1$ or $-1$ ($u^2-v^2=1$ or $-1$) for real (pure imaginary) $E$.
See Appendix for the details.

\subsubsection{Real-positive eigenvalue state}
For a real positive $E$, there are two propagating modes 
and two damping/growing modes
[Figs.~\ref{fig:Energy2_ep_binary}(a) and \ref{fig:Energy2_ep_binary}(c)] respectively given by
\begin{align}
    \begin{pmatrix}a_r \\ -b_r \end{pmatrix}e^{\pm i k_1 x},
\end{align}
and
\begin{align}
    \begin{pmatrix} b_r \\ a_r \end{pmatrix}e^{\mp q_2 x},
\end{align}
where
\begin{align}
a_r&=\mbox{sgn}(\beta)\sqrt{\frac{\sqrt{\beta^2 + E^2}}{2E}+\frac{1}{2}}, \label{eq:def_ar}\\
		b_r&=\sqrt{\frac{\sqrt{\beta^2 + E^2}}{2E}-\frac{1}{2}}, \label{eq:def_br}\\
		k_1&=\sqrt{2}\sqrt{\sqrt{\beta^2+E^2} - \beta},\\
		q_2&=\sqrt{2}\sqrt{\sqrt{\beta^2+E^2} + \beta},
\end{align}
and $a_r$ and $b_r$ satisfies $|a_r|^2-|b_r|^2=1$.
These are eigensolutions in the whole region of $\beta$: $-1\le\beta\le 1$. 
\subsubsection{Pure-imaginary eigenvalue state}
\begin{figure}
	\centering
    \includegraphics[clip,width=8.5cm]{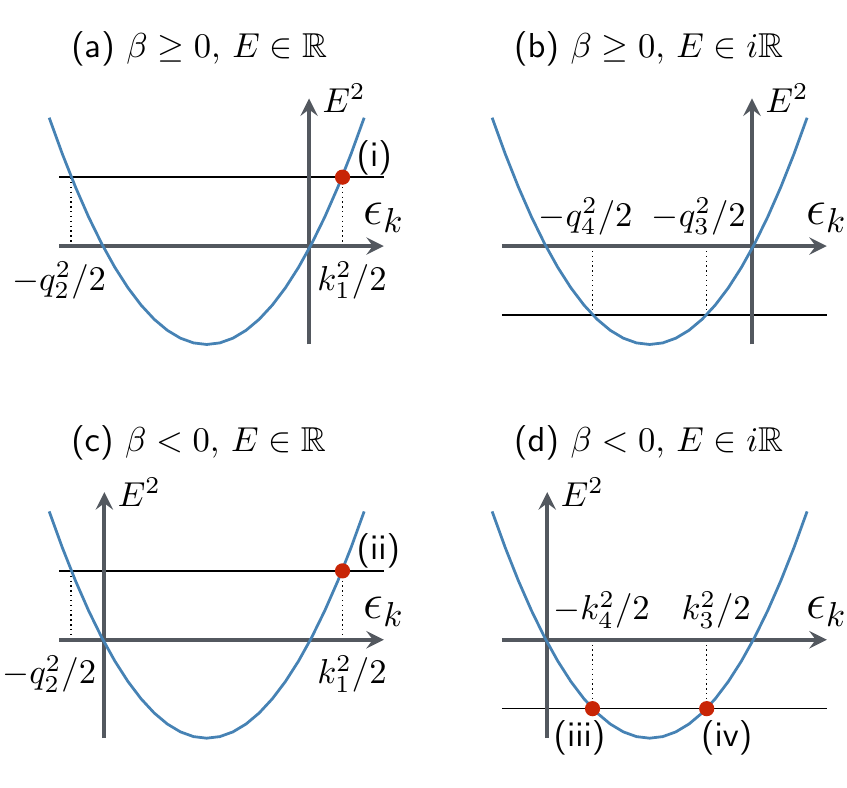}	
		\caption{
		Spin-wave spectrum [Eq.~\eqref{eq:Bdg_binary_magnon_eigenvalue}] for a binary BEC. Shown are $E^2$ as a function of $\epsilon_k=k^2/2$ for $\beta\ge 0$ (a),(b) and $\beta<0$ (c),(d).
		We solve $k$ for a given $E^2$, where $E^2>0$ ($E^2<0$) means a real (pure imaginary) eigenvalue $E$. The solutions with $\epsilon_k>0$ ($\epsilon_k<0$) are propagating (localized) modes. 
		The points indicated by (i)--(iv) correspond to the solutions of the incident wave for the cases of (i)--(iv) with the asymptotic forms given by Eqs.~\eqref{eq:binary_BC(a)}--\eqref{eq:binary_BC(d)}.}
	\label{fig:Energy2_ep_binary}
\end{figure}

We rewrite the pure imaginary eigenvalue $E$ as $E=i\Delta$, where $\Delta\in \mathbb{R}$ and $|\Delta|<|\beta|$.
For $\beta < 0$, there are four propagating solutions [Fig.~\ref{fig:Energy2_ep_binary}(d)]:
\begin{align}
    \begin{pmatrix} b_c \\ a_c\end{pmatrix} e^{\pm ik_3 x}\ \ \textrm{and}\ \ 
    \begin{pmatrix} a_c \\ -b_c\end{pmatrix} e^{\pm ik_4 x},
\label{eq:binary_pw_imE}
\end{align}
which change to four damping/growing solutions for $\beta\ge 0$ [Fig.~\ref{fig:Energy2_ep_binary}(b)]:
\begin{align}
    \begin{pmatrix} b_c \\ a_c\end{pmatrix} e^{\mp q_3 x}\ \ \textrm{and}\ \ 
    \begin{pmatrix} a_c \\ -b_c\end{pmatrix} e^{\mp q_4 x},
\end{align}
where 
\begin{align}
    a_c&={\rm sgn}(\beta)\frac{e^{-i\theta/2}}{\sqrt{-2i\sin\theta}},\ \ 
    b_c=\frac{e^{i\theta/2}}{\sqrt{-2i\sin\theta}}
\label{eq:def_ac_bc}
\end{align}
with $\theta\equiv \sin^{-1}(\Delta/|\beta|)$, and
\begin{align}
    k_3=iq_3&=\sqrt{2}\sqrt{+\sqrt{\beta^2-\Delta^2} - \beta},\\
    k_4=iq_4&=\sqrt{2}\sqrt{-\sqrt{\beta^2-\Delta^2} - \beta}.
\end{align}
Here, $a_c$ and $b_c$ are normalized as $a_c^2-b_c^2=1$.
We define $k_{3,4}$ and $q_{3,4}$ to be positive real value and do not use when they are imaginary.


\subsection{Tunneling properties of the spin-wave modes\label{sec:tunneling properties_magnon}}

We now solve the BdG Eq.~\eqref{eq:BdG_binary_magnon_x}
and calculate the reflection and transmission probabilities of quasiparticles with energy $E$ and momentum $k_{\rm in}$ injected from the left.
As for the barrier potential $U(x)$, 
we use the Gaussian potential given by
\begin{align}
    \label{eq:potential}
    U(x) = U_0e^{-x^2/(2\sigma^2)}.
\end{align}
We numerically solve the BdG Eq.~\eqref{eq:BdG_binary_magnon_x} by means of the finite element method with imposing the asymptotic form of the wave function at $x\to\pm\infty$.

For the case of $\beta\ge 0$, a quasiparticle with $k_{\rm in}=k_1$ injected from the left ($e^{ik_1 x}$) is reflected to the left  ($e^{-ik_1 x}$) or transmitted to the right  ($e^{ik_1 x}$).
In addition, localized modes at the potential barrier appear ($e^{\pm q_2 x}$). 
The asymptotic form of the wave function at $x\to \pm \infty$ is then given by
\begin{description}
\item [(i)]  $\beta\ge 0, E\in \mathbb{R}$ with an incident wave $e^{ik_1 x}$:
\begin{align}
    &\begin{cases}
	\begin{pmatrix}
	u_{-\infty} \\ v_{-\infty}
	\end{pmatrix} 
	=
	\begin{pmatrix}
		a_r \\  -b_r
	\end{pmatrix} 
	(e^{ik_1x} + Re^{-ik_1x})
	+A
	\begin{pmatrix}
		b_r \\  a_r
	\end{pmatrix}
	e^{q_2x},\ \ \\
	\begin{pmatrix}
	u_{+\infty} \\ v_{+\infty}
	\end{pmatrix} 
	=T
	\begin{pmatrix}
		a_r \\ -b_r
	\end{pmatrix} 
	e^{ik_1x} + B
	\begin{pmatrix}
		b_r \\ a_r
	\end{pmatrix}
	e^{-q_2x}.
    \end{cases}
\label{eq:binary_BC(a)}
\end{align}
\end{description}
Here, $R$ and $T$ are the reflection and transmission coefficients, respectively, and $A$ and $B$ are the coefficients for the localized modes.
The current conservation law requires $|R|^2+|T|^2=1$.

For the case of $\beta<0$, there are three possibilities:
(ii) real positive $E\in \mathbb{R}$ with $k_{\rm in}=k_1$,
(iii) pure imaginary $E=i\Delta\in i\mathbb{R}$ ($\Delta>0$) with $k_{\rm in}=-k_4$,
and (iv) pure imaginary $E=i\Delta\in i\mathbb{R}$ with $k_{\rm in}=k_3$.
The results for $\Delta<0$ can be obtained by taking the complex conjugate of the wave functions obtained for $\Delta>0$.
The points corresponding to the incident wave are depicted in the dispersion relation in Fig.~\ref{fig:Energy2_ep_binary}.
Note that though the group velocity of a quasiparticle with a pure-imaginary eigenvalue is zero, the causality is satisfied when we assume $d\Delta/dk$ to be a group velocity and choose the momentum of the incident, reflected, and transmitted waves.
Hence, the asymptotic forms of the wave function at $x\to \pm \infty$ are respectively given as follows:
\begin{description}
\item[(ii)]  $\beta<0, E\in \mathbb{R}$  with $k_{\rm in}=k_1$:\\The asymptotic form is the same as Eq.~\eqref{eq:binary_BC(a)}.

\item [(iii)] $\beta<0, E\in i\mathbb{R}$  with $k_{\rm in}=-k_4$:
\begin{align}
&\begin{cases}
	\begin{pmatrix}
	u_{-\infty} \\ v_{-\infty}
	\end{pmatrix} 
	=
	\begin{pmatrix}
		a_c \\  -b_c
	\end{pmatrix} 
	(e^{-ik_4x} + Re^{ik_4x})
	+A
	\begin{pmatrix}
		b_c \\  a_c
	\end{pmatrix}
	e^{-ik_3x},\\
	\begin{pmatrix}
	u_{+\infty} \\ v_{+\infty}
	\end{pmatrix} 
	= T
	\begin{pmatrix}
		a_c \\ -b_c
	\end{pmatrix} 
	e^{-ik_4x} + B
	\begin{pmatrix}
		b_c \\ a_c
	\end{pmatrix}
	e^{ik_3x}.
\end{cases}
\label{eq:binary_BC(c)}
\end{align}
\item [(iv)] $\beta<0, E\in i\mathbb{R}$  with $k_{\rm in}=k_3$:
\begin{align}
&\begin{cases}
	\begin{pmatrix}
	u_{-\infty} \\ v_{-\infty}
	\end{pmatrix} 
	=
	\begin{pmatrix}
		b_c \\  a_c
	\end{pmatrix} 
	(e^{ik_3x} + Re^{-ik_3x})
	+A
	\begin{pmatrix}
		a_c \\  -b_c
	\end{pmatrix}
	e^{ik_4x},\\
	\begin{pmatrix}
	u_{+\infty} \\ v_{+\infty}
	\end{pmatrix} 
	= T
	\begin{pmatrix}
		b_c \\ a_c
	\end{pmatrix} 
	e^{ik_3x} + B
	\begin{pmatrix}
		a_c \\ -b_c
	\end{pmatrix}
	e^{-ik_4x}.
\end{cases}
\label{eq:binary_BC(d)}
\end{align}
\end{description}
Here, we further introduce the scaled coefficients
\begin{align}
    \tilde{A}=\sqrt{-\frac{k_{\rm sc}}{k_{\rm in}}} A,\ \ 
    \tilde{B}=\sqrt{-\frac{k_{\rm sc}}{k_{\rm in}}} B,
\end{align}
where ${k}_{\rm sc}$ is the real part of the momentum of the $A$ term. For example, $k_{\rm sc}=0, -k_3$ and $k_4$ for the cases (ii), (iii), and (iv), respectively.
Using $\tilde{A}$ and $\tilde{B}$,
the current conservation law is written as 
\begin{align}
    |T|^2+|R|^2+|\tilde{A}|^2+|\tilde{B}|^2=1
    \label{eq:current_conservation}
\end{align} for all cases of (i)--(iv).

\begin{figure}
	\centering
    \includegraphics[clip,width=8.5cm]{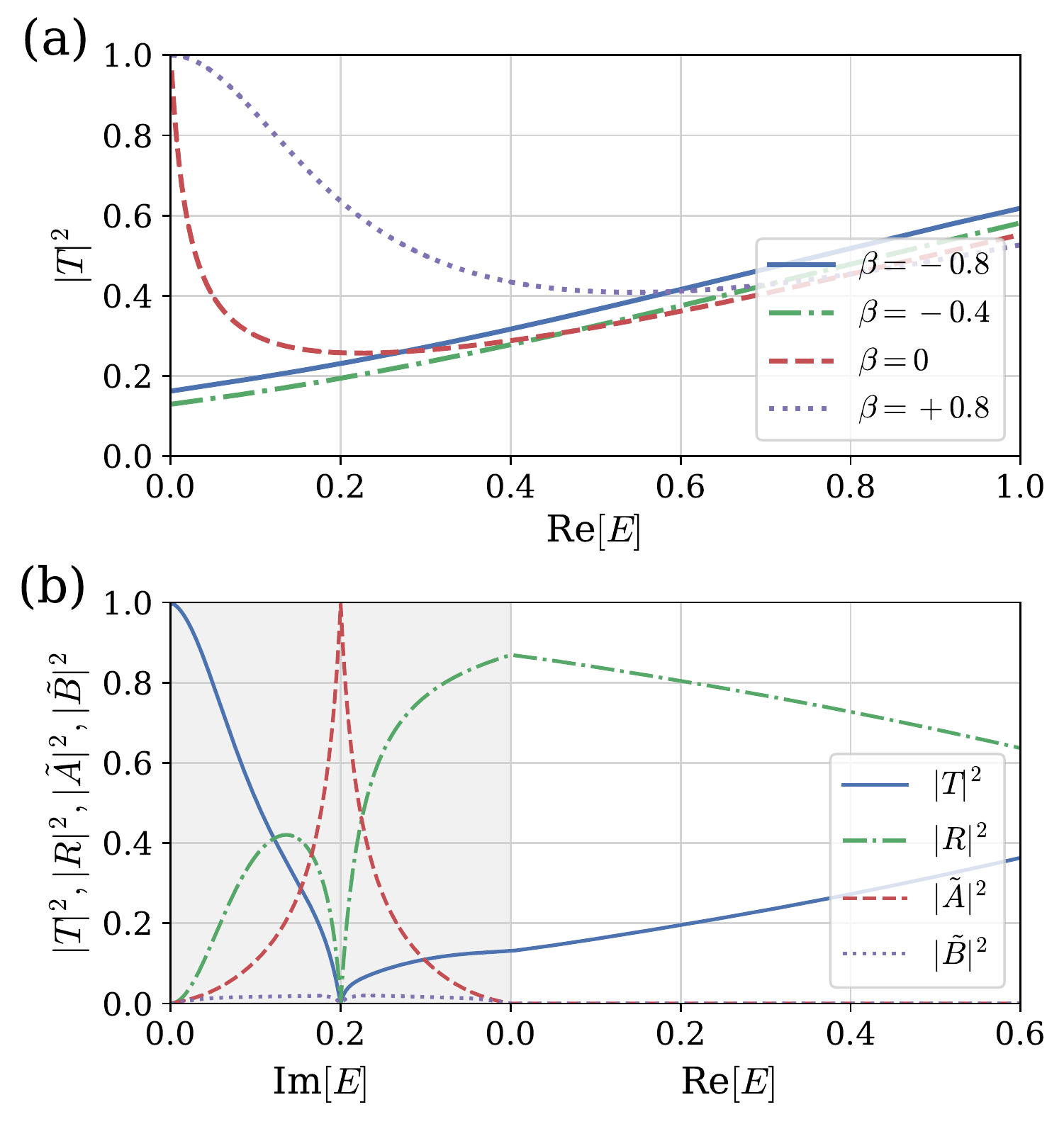}	
		\caption{Tunneling property of spin-wave modes in a binary BEC. (a) Transmission probability $|T|^2$ for real $E$, obtained by imposing the asymptotic forms of Eq.~\eqref{eq:binary_BC(a)} [cases (i) and (ii)]. 
        The perfect transmission occurs in the low-energy limit, i.e. $|T|^2\rightarrow1$ as $E\rightarrow0$, for $\beta\ge 0$, whereas $|T|^2$ at $E\to 0$ for $\beta<0$ is smaller than unity.
		(b) $E$ dependence of the tunneling coefficients $|T|^2, |R|^2, |\tilde{A}|^2$, and $|\tilde{B}|^2$ in the presence of dynamical instability at $\beta=-0.2$. The horizontal axis is chosen such that $\epsilon_{k_{\rm in}}=0$ at the left edge of the panel and $\epsilon_{k_{\rm in}}$ increases as one goes to the right. Correspondingly, along the horizontal axis, the value of $E$ starts from 0, ${\rm Im}E$ first increases [case (iii)], takes maximum value $0.2$, and decreases to $0$ [case (iv)], and then $E$ changes to real [case (ii)].
		The region where $E$ is pure imaginary is shaded with gray.
		The perfect transmission ($|T|^2=1$) occurs at $E=0$ at the left edge of the panel, and the perfect reflection ($|\tilde{A}|^2=1$) occurs at $E=0.2i$.
		We use the barrier potential  with $U_0=2$ and $\sigma=0.5$ for both panels. The perfect transmission and the perfect reflection occurs independently from the values of $U_0$ and $\sigma$.
        }
	\label{fig:binary_coeff}
\end{figure}

\begin{figure}
	\centering
    \includegraphics[clip,width=8.5cm]{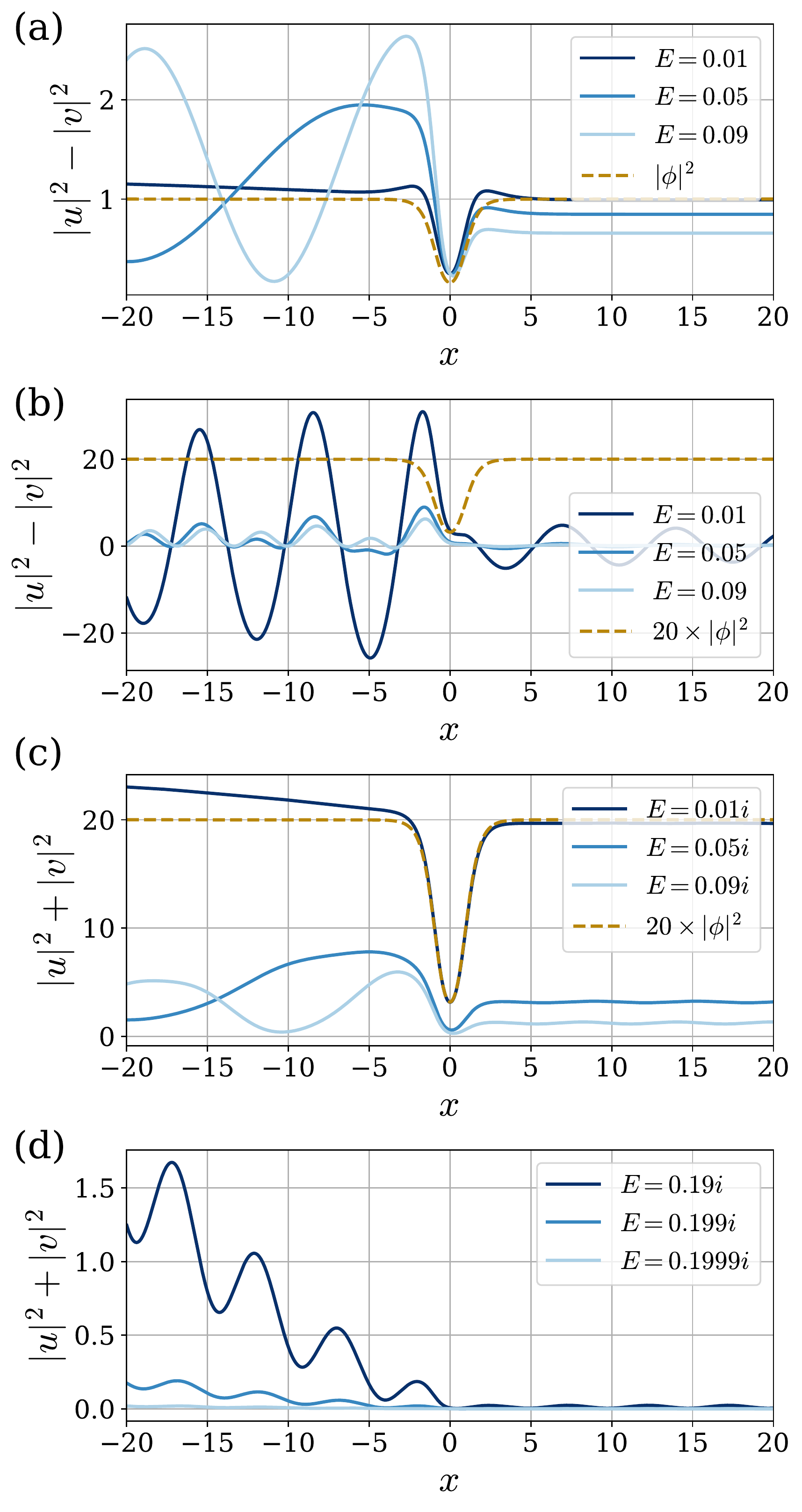}	
		\caption{Quasiparticle wave function of spin-wave modes in a binary BEC. Shown are $|u|^2 - |v|^2$ for real eigenvalue modes (a)(b), and $|u|^2+|v|^2$ for pure-imaginary eigenvalue modes (c)(d).
		The panels (a) and (b) are the results for the zero-energy limit at $\beta=0.2$ and $-0.2$, respectively, obtained for the asymptotic form of Eq.~\eqref{eq:binary_BC(a)} [cases (i) and (ii)]. As $E$ approaches to 0, $|u|^2-|v|^2$ approaches to $|\phi|^2$ in (a), whereas its $x$ dependence has a distinctive difference from $|\phi|^2$ in (b), being consistent with the presence (a) and absence (b) of the perfect transmission (see text).
		Panels (c) and (d) are the results for $\beta=-0.2$ with the asymptotic form of Eq.~\eqref{eq:binary_BC(c)} [case (iii)] in the limit of $E\to 0$ and $E\to i|\beta|$, respectively.
		As $E$ approaches to 0, $|u|^2+|v|^2$ becomes proportional to $|\phi|^2$ in (c), being consistent with the conditions for the perfect transmission.
		In (d), the quasiparticle wave function becomes smaller as $E$ approaches to $i|\beta|$. This is due to the destructive interference between the incident and reflected waves.
        For all panels, we use the barrier potential  with $U_0=2$ and $\sigma=0.5$.
    	} 
	\label{fig:binary_wf_J}
\end{figure}

In Fig.~\ref{fig:binary_coeff}, we show the numerical results for the tunneling coefficients for the cases (i)--(iv). The results for the quasiparticle wave functions are depicted in Fig.~\ref{fig:binary_wf_J}. Figure~\ref{fig:binary_coeff}(a) shows the transmission probability for real $E$ ($>0$) with various values of $\beta$, obtained by imposing the asymptotic form of Eq.~\eqref{eq:binary_BC(a)} [cases (i) and (ii)]. One can see that the perfect transmission occurs at the low energy limit for $\beta\ge 0$, i.e., $|T|^2$ goes to unity as $E\to 0$,
being consistent with the previous works~\cite{PhysRevA.83.053624,PhysRevA.84.013616}.
On the other hand, for $\beta<0$, the transmission probability becomes smaller than unity.
This difference is understood as follows.
It is known that the perfect transmission occurs when the quasiparticle wave function coincides with the condensate wave function: $u(x)=-v^*(x)=\phi(x)$.
In the present case,
the incident momentum $k_1$ goes to zero (nonzero) as $E\to 0$ when $\beta\ge 0$ ($\beta<0$), and hence $u(x)=-v^*(x)=\phi(x)$ is (is not) satisfied at $E\to 0$.
In Fig.~\ref{fig:binary_wf_J}(a) and \ref{fig:binary_wf_J}(b), 
we show $|u|^2-|v|^2$ at $\beta=0.2$ and $-0.2$, respectively.
[The integral $\int (|u|^2-|v|^2) dx$ gives the norm of the quasiparticle wave function for a real $E$, whereas it should be vanish for a complex $E$. We therefore plot $|u|^2+|v|^2$ for $\beta<0$ as shown in Figs.~\ref{fig:binary_wf_J}(c) and \ref{fig:binary_wf_J}(d). See also Appendix.]
One can see from these figures that $|u|^2-|v|^2$ for $\beta=0.2$ becomes close to $|\phi|^2$ as $E\to 0$, whereas it for $\beta=-0.2$ at $E\to 0$ has a completely different $x$ dependence from $|\phi|^2$,
being consistent with the above discussion.
The result in the previous works~\cite{Danshita_2006,PhysRevA.78.043601,doi:10.1143/JPSJ.78.023001} that the phonon mode in a scalar BEC moving with the critical current
does not exhibit the perfect transmission originates from the same reason, where $u$ and $v^*$ deviate from $\phi$ due to local enhancement of density fluctuations around the potential barrier~\cite{doi:10.1143/JPSJ.78.023001}.

We note that
$k_4$ goes to zero and $u(x)=-v^*(x)=\phi(x)$ is satisfied when $E$ goes to zero along the imaginary axis.
Hence, the perfect transmission at $E\to0$ occurs in this case.
Figure~\ref{fig:binary_coeff}(b) shows the $E$ dependence of $|T|^2$, $|R|^2$, $|\tilde{A}|^2$ and $|\tilde{B}|^2$ at $\beta=-0.2$.
Here, we set the horizontal axis of Fig.~\ref{fig:binary_coeff}(b) such that $\epsilon_{k_{\rm in}}$ is zero at the left end of the figure and $\epsilon_{k_{\rm in}}$ increases as one goes right.
The region where $E$ is pure imaginary is shaded with gray.
We numerically confirmed the current conservation law given by Eq.~\eqref{eq:current_conservation}.
In Fig.~\ref{fig:binary_coeff}(b), the transmission probability $|T|^2$ goes to unity as ${\rm Im}\,E\to 0$ at the left end of the figure, indicating that the perfect transmission occurs even for the dynamically unstable modes.
We also confirm that $|u|^2+|v|^2$ shown in Fig.~\ref{fig:binary_wf_J}(c) becomes proportional to $|\phi|^2$ as ${\rm Im}E\to 0$, being consistent with $u(x)=-v^*(x)=\phi(x)$.
On the other hand, at the point where $E$ changes from pure imaginary to real values [the point $E=0$ and $\epsilon_k>0$ in Fig.~\ref{fig:Energy2_ep_binary}(c) and \ref{fig:Energy2_ep_binary}(d), which corresponds to the boundary between the gray shaded region and the white region in Fig.~\ref{fig:binary_coeff}(b)], the incident momentum $k_3$ remains nonzero in the limit of $E\to 0$, which means $u(x)=-v^*(x)=\phi(x)$ is not satisfied in this limit,
resulting in the absence of the perfect transmission.

We also find from Figs.~\ref{fig:binary_coeff}(b) that $|\tilde{A}|^2$ goes to unity at $E=i|\beta|$. At this point, $|{\rm Im}E|$ takes its maximum value, i.e., $E\in i\mathbb{R}$ and $d{\rm Im}E/dk=0$. When $E$ is pure imaginary, the $A$ term in Eqs.~\eqref{eq:binary_BC(c)} and \eqref{eq:binary_BC(d)} represents the reflected wave with the momentum different from incident one.
Hence, $|\tilde{A}|^2=1$ means occurrence of perfect reflection.
The origin of the perfect reflection is because the number of linearly independent solutions given in Eq.~\eqref{eq:binary_pw_imE} decreases at $k_3=k_4$
and the solution that satisfies the asymptotic forms of Eqs.~\eqref{eq:binary_BC(c)} and \eqref{eq:binary_BC(d)} disappear.
Hence, the incident wave and the reflected wave (the $A$ term) destructively interfere.
The destructive interference can be confirmed in the quasiparticle wave function shown in Fig.~\ref{fig:binary_wf_J}(d), in which we plot $|u|^2+|v|^2$ obtained for $\beta=-0.2$ with the asymptotic form of Eq.~\eqref{eq:binary_BC(c)}. The wave function goes to zero as $E\to i|\beta|$ even in the $x<0$ region.
Since we fix the amplitude of the incident wave, the reduction of $|u|^2+|v|^2$ means the destructive interference.
Perfect reflection 
was also predicted to occur for the long-wave length limit of the spin-wave mode in a ferromagnetic BEC~\cite{PhysRevA.83.053624,PhysRevA.84.013616}.
The perfect reflection we find in this paper is different from that in the previous works in that the origin for the perfect reflection in the latter case is essentially the same as the ordinary quantum-mechanical tunneling of a free particle.

\section{Spin-1 polar BEC\label{sec:spin1_BEC}}
\subsection{Model\label{sec:model_spin1-BEC}}
Next, we consider the tunneling problem in a spin-1 polar BEC.
The BdG equation for the spin-wave mode in this system is the same as the one in the previous section
except for the quadratic Zeeman energy $q_z$ term.
Since the quadratic Zeeman effect breaks the spin rotational symmetry,
the spin-wave spectrum has a nonzero eigenvalue at $k=0$.
Here, the eigenvalue at $k=0$ can be real or pure imaginary depending on the value of $q_z$.
We therefore focus on the $q_z$ dependence of the tunneling properties.
Below, we redefine the variables used in the previous section, so that the resulting BdG equation has the same form as in the previous section except for the $q_z$ term.

The energy functional of a spin-1 system is given by
\begin{align}\label{eq:Energy_func_spin1BEC}
    \mathcal{E}_\textrm{spin-1}=&\int d\bm x\sum_{m}\left\{\frac{\hbar^2}{2M}\left|\frac{\partial \Psi_m}{\partial x}\right|^2+\left[ U(x)+q_z m^2\right]|\Psi_m|^2\right\}\nonumber\\
      &+\frac{1}{2}\int d\bm x\left[c_0n^2+c_1|\bm{F}|^2\right],
\end{align}
where $\Psi_m$ is the condensate wave function of the atoms in the magnetic sublevel $m=1,0$, and $-1$, $M$ is atomic mass, $U(x)$ is barrier potential, $n(x)=\sum_m|\psi_m(x)|^2$ is condensate density, and $\bm{F}(x) =(F_x(x),F_y(x),F_z(x))$ is the spin density vector defined by $\bm{F}(x)=\sum_{mm'}\Psi_{m}^*(x)\bm{S}_{mm'}\Psi_{m'}(x)$ with $\bm{S}=(S_x,S_y,S_z)$ being the spin-1 matrices.
The interaction coefficients are given by $c_0=4\pi\hbar^2(2a_2+a_0)/(3M)$ and $c_1=4\pi\hbar^2(a_2-a_0)/(3M)$, where $a_{\mathcal{F}}$ is the $s$-wave scattering length for the total spin $\mathcal{F}=0,2$ channel. 

The ground-state phase of this system is determined by the values of $c_1$ and $q_z$. (The phase diagram is given, e.g., in Ref.~\citen{PhysRevA.87.061604}.)
Below, we consider a condensate in the $m=0$ state, i.e., a polar BEC.
The polar BEC is the ground state of the system when
$q_z>{\rm max}(0,-2c_1n)$,
and the system becomes dynamically unstable when $(c_1,q_z)$ is outside of this region.

The GP and BdG equations are obtained by the same manner as the previous section.
For the case of a polar BEC,
the stationary solution $(\Psi_1,\Psi_0,\Psi_{-1})=(0,\Phi(x),0)$ satisfies
\begin{align}
    \mu \Phi=\left[-\frac{\hbar^2}{2M}\frac{d^2}{dx^2}+U(x)+c_0|\Phi|^2\right]\Phi,
    \label{eq:GP_spin1BEC_stationary}
\end{align}
where $\mu$ is the chemical potential. 
Assuming $U(x)\rightarrow0$ and $|\Phi|^2\rightarrow n_0 (=\textrm{const.})$ at $x\rightarrow\pm\infty$, we obtain $\mu=c_0n_0$.
We use this $\mu$ to scale the dimensionful variables, i.e., we rescale the energy, length, and time scales by $\mu$, $\hbar/\sqrt{M\mu}$, and $\mu/\hbar$.
Rewriting $\Phi(x)=\sqrt{n_0}\phi(x)$, Eq.~\eqref{eq:GP_spin1BEC_stationary} reduces to
\begin{align}
    &\left(\mathcal{L}_0-1+|\phi|^2\right)\phi=0,
    \label{eq:GP_spin1BEC_stationary2}
\end{align}
where $\mathcal{L}_0$ is defined in Eq.~\eqref{eq:def_L0}.
Equation~\eqref{eq:GP_spin1BEC_stationary2} is identical to Eq.~\eqref{eq:GP_binaryBEC_stationary2}.

We introduce the interaction parameter $\beta$ as
\begin{align}
    \beta \equiv \frac{c_1}{c_0}.
\end{align}
Rewriting the wave function as $\Psi_{m}(x)=\sqrt{n_0}\psi_{m}(x)$,
the time-dependent GP equation in the dimensionless form is given by
\begin{subequations}
\begin{align}
    i\frac{\partial\psi_{\pm 1}}{\partial t} 
    =&\left[\mathcal{L}_0+1\pm\beta f_z+q_z\right]\psi_{\pm 1}+\frac{\beta}{\sqrt{2}}f_{\mp}\psi_{0},\\
    i\frac{\partial\psi_0}{\partial t} 
    =&\left[\mathcal{L}_0+1\right]\psi_0 +\frac{\beta}{\sqrt{2}}(f_{+}\psi_{+1}+f_{-}\psi_{-1}),
\end{align}
\label{eq:GP_spin1BEC2}
\end{subequations}
where $f_+=f_-^*=\sqrt{2}(\psi_1^*\psi_0+\psi_0^*\psi_{-1})$ and $f_z=|\psi_1|^2-|\psi_{-1}|^2$.
Substituting
\begin{align}
    \begin{pmatrix}
    \psi_{+1}\\ \psi_{0} \\ \psi_{-1}
    \end{pmatrix}
    = e^{-it}\left[
    \begin{pmatrix}
    0\\ \phi \\ 0
    \end{pmatrix}
    +\bm{u}e^{-iEt/\hbar}+\bm{v}^*e^{+iE^*t/\hbar}
     \right]
\end{align}
into Eq.~\eqref{eq:GP_spin1BEC2} and linearizing the equation with respect to $\bm{u}$ and $\bm{v}$, with $\bm{u}=(u_{+1},u_0,u_{-1})^{\rm T}$ and $\bm{v}=(v_{+1},v_0,v_{-1})^{\rm T}$ being three-component spinors, 
the BdG equation for a polar BEC is given by
\begin{align}\label{eq:bdg_spin1}
  \begin{pmatrix}
    H_0 + H_1  & H_2         \\
    -H^*_2                          & -[H_0 + H^*_1] 
  \end{pmatrix}
  \begin{pmatrix}
    \bm{u} \\ \bm{v}
  \end{pmatrix}
  =E
  \begin{pmatrix}
    \bm{u} \\ \bm{v}
  \end{pmatrix}
\end{align}
where
\begin{subequations}
    \begin{align}
        H_0&=
        \begin{pmatrix}
            \mathcal{L}_0-1+q_z & 0               & 0            \\
            0                   & \mathcal{L}_0-1 & 0            \\
            0                   & 0               & \mathcal{L}_0-1+q_z
        \end{pmatrix},\\    
        H_1&=|\phi|^2
        \begin{pmatrix}
            1+\beta & 0 & 0            \\
            0           & 2 & 0            \\
            0           & 0 & 1+\beta
        \end{pmatrix},\\
        H_2&=\phi^2
        \begin{pmatrix}
            0     & 0 & \beta \\
            0     & 1 & 0     \\
            \beta & 0 & 0
        \end{pmatrix}.
    \end{align}    
\end{subequations}
This $6\times 6$ eigenvalue equation can be divided into three $2\times 2$ equations: 
Defining the density-wave mode $u^{\rm{d}}=u_0,v^{\rm{d}}=v_0$ and spin-wave modes $u_\pm^{\rm{s}}=u_{\pm 1},v_\pm^{\rm{s}}=v_{\mp 1}$, the BdG Eq.~\eqref{eq:bdg_spin1} reduces to
\begin{widetext}
\begin{align}
    \begin{pmatrix}
    \mathcal{L}_0 -1 + 2|\phi|^2 & \phi^2 \\
    -(\phi^*)^2 & -(\mathcal{L}_0-1 +2|\phi|^2)
    \end{pmatrix}
    \begin{pmatrix} u^{\rm d} \\ v^{\rm d} \end{pmatrix}
    &=E^{\rm d}\begin{pmatrix} u^{\rm d} \\ v^{\rm d} \end{pmatrix},
    \label{eq:BdG_spin1_phonon_x}\\
    \begin{pmatrix}
    \mathcal{L}_0 -1 +q_z+ (1+\beta)|\phi|^2 & \beta\phi^2 \\
    -\beta(\phi^*)^2 & -(\mathcal{L}_0 -1 +q_z+(1+\beta)|\phi|^2)
    \end{pmatrix}
    \begin{pmatrix} u_\pm^{\rm s} \\ v_\pm^{\rm s} \end{pmatrix}
    &=E^{\rm s}\begin{pmatrix} u_\pm^{\rm s} \\ v_\pm^{\rm s} \end{pmatrix}.
    \label{eq:BdG_spin1_magnon_x}
\end{align}
\end{widetext}
Equation~\eqref{eq:BdG_spin1_phonon_x} has a zero-energy solution $u^{\rm d}=-(v^{\rm d})^*=\phi$, indicating that the density-wave is the NG phonon associated with the spontaneous breaking of the U(1) gauge symmetry.
On the other hand, the spin-wave mode is not NG mode for $q_z\neq 0$,
since the SO(3) spin rotational symmetry which is broken in the presence of the quadratic Zeeman energy.
The previous works investigated the tunneling properties for the BdG Eq.~\eqref{eq:BdG_spin1_magnon_x} with $q_z=0$ and $\beta>0$ and showed that the perfect transmission occurs in the low energy limit~\cite{PhysRevA.83.053624,PhysRevA.84.013616}.

Note that Eq.~\eqref{eq:BdG_spin1_magnon_x} at $q_z=0$ is identical to Eq.~\eqref{eq:BdG_binary_magnon_x}.
The role of the $q_z$ is that it effectively shifts the chemical potential from 1 to $1-q_z$,
opening an energy gap at $k=0$ in a uniform system (see below).
This is possible because the condensation occurs in the different internal state from the quasiparticles.

\subsection{Bogoliubov spectrum in a uniform system\label{sec:Bogoliubov_spectrum_spin1BEC}}
We analytically solve the BdG Eq.~\eqref{eq:BdG_spin1_magnon_x}
for $U(x)=0$ and obtain propagating and damping/growing solutions.
Substituting $(u^{\rm s},v^{\rm s})^{\rm T}=(ue^{ikx},ve^{ikx})^{\rm T}$ and $\phi=1$ to Eq.~\eqref{eq:BdG_spin1_magnon_x}, we obtain
\begin{align}\label{eq:bdg_polar}
  \begin{pmatrix}
    \epsilon_k + q_z + \beta & \beta                       \\
    -\beta           		& -(\epsilon_k + q_z + \beta)
  \end{pmatrix}
  \begin{pmatrix}
      u\\v
  \end{pmatrix}  
    =E^{\rm{s}}
  \begin{pmatrix}
    u\\v
  \end{pmatrix},
\end{align}
which has the eigenvalue
\begin{align}\label{eq:bdg_polar_eigenvalue}
    E^{\rm{s}}=\sqrt{(\epsilon_k+q_z)(\epsilon_k+q_z+2\beta)}.
\end{align}
It follows that when $q_z<{\rm max}(0,-2\beta)$,
$E^{\rm s}$ becomes pure imaginary for a certain region of $k$ and the system becomes dynamically unstable.
The condition for the dynamical stability $q_z>{\rm max}(0,-2\beta)$ agrees with the region for which the polar state is the ground state.

\subsubsection{Real-positive eigenvalue state}
For a real positive $E$, the four linearly independent solutions are given by
\begin{align}
    \begin{pmatrix}a_r \\ -b_r \end{pmatrix}e^{\pm i k_1 x}\ \textrm{and}\ 
    \begin{pmatrix} b_r \\ a_r \end{pmatrix}e^{\mp ik_2 x},
    \label{eq:spin1BEC_PW_real}
\end{align}
where $a_r$ and $b_r$ are defined in Eqs.~\eqref{eq:def_ar} and \eqref{eq:def_br}, respectively, and 
$k_1$ and $k_2$ are given by
\begin{align}
	k_1&=\sqrt{2}\sqrt{+\sqrt{\beta^2+E^2}-(\beta+q_z)}, \label{eq:spin1_k1}\\
	k_2&=\sqrt{2}\sqrt{-\sqrt{\beta^2+E^2}-(\beta+q_z)}. \label{eq:spin1_k2}
\end{align}
The right-hand sides of Eqs.~\eqref{eq:spin1_k1} and \eqref{eq:spin1_k2} can be real or pure imaginary.
When $k_{1,2}$ is pure imaginary, we rewrite it as $k_{1,2}=iq_{1,2}$ and use real-valued $q_{1,2}$($>0$).
The corresponding solutions in Eq.~\eqref{eq:spin1BEC_PW_real} express both
propagating modes ($e^{\pm i k_{1,2}x}$) and growing/damping modes ($e^{\mp q_{1,2}x}$).

\subsubsection{Pure-imaginary eigenvalue state}
Rewriting the eigenvalue as $E=i\Delta$ with $\Delta\in \mathbb{R}$ and $|\Delta|<|\beta|$,
the four solutions are given by:
\begin{align}
    \begin{pmatrix} b_c \\ a_c\end{pmatrix} e^{\pm ik_3 x}\ \ \textrm{and}\ \ 
    \begin{pmatrix} a_c \\ -b_c\end{pmatrix} e^{\pm ik_4 x},
\end{align}
where $a_c$ and $b_c$ are defined in Eq.~\eqref{eq:def_ac_bc} and
\begin{align}
    k_3&=\sqrt{2}\sqrt{+\sqrt{\beta^2-\Delta^2} - (\beta+q_z)}, \label{eq:spin1_k3}\\
    k_4&=\sqrt{2}\sqrt{-\sqrt{\beta^2-\Delta^2} - (\beta+q_z)}.\label{eq:spin1_k4}
\end{align}
The most right-hand sides of Eqs.~\eqref{eq:spin1_k3} and \eqref{eq:spin1_k4} can be real or pure imaginary.
When $k_{3,4}$ is pure imaginary, we rewrite it as $k_{3,4}=iq_{3,4}$ and use real-valued $q_{3,4}$ ($>0$).
The corresponding solutions in Eq.~\eqref{eq:spin1BEC_PW_real} express both
propagating modes ($e^{\pm i k_{3,4}x}$) and growing/damping modes ($e^{\mp q_{3,4}x}$).

\begin{figure}
	\centering
    \includegraphics[clip,width=8.5cm]{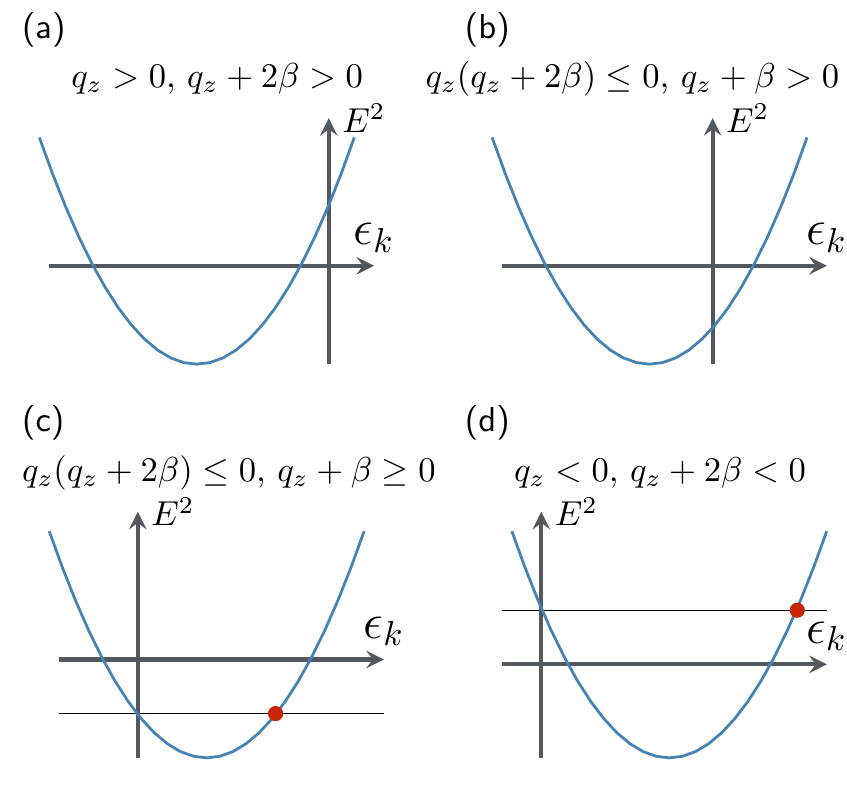}	
		\caption{
		Spin-wave spectrum [Eq.~\eqref{eq:bdg_polar_eigenvalue}] for a polar BEC. Shown are $E^2$ as the function $\epsilon_k$ for  (a)$q_z>0$ and $q_z+2\beta>0$, (b)$q_z(q_z+2\beta)\le 0$ and $q_z+\beta>0$, (c)$q_z(q_z+2\beta)\le 0$ and $q_z+\beta\ge 0$, and (d)$q_z<0$ and $q_z+2\beta<0$.
		We solve $k$ for a given $E^2$, where $E^2>0$ $(E^2<0)$ means a real (pure
		imaginary) eigenvalue $E$. Red circles in (c) and (d) indicate the points $E = \sqrt{q_z(q_z+2\beta)}$, where the transition probability resonantly increases.
		}
	\label{fig:Energy2_ep_polar}
\end{figure}

\subsection{Tunneling properties of the spin-wave modes\label{sec:tunneling_properties_spin1BEC}}

We now investigate the tunneling properties of the spin-wave modes in a polar BEC.
In the same way as the previous section, for given energy $E$ and momentum  $k_{\rm in}$ of the incident wave, we construct the asymptotic form of the quasiparticle wave function at $x\to\pm\infty$ and numerically solve the BdG Eq.~\eqref{eq:BdG_spin1_magnon_x} using the finite element method.

\begin{figure}[H]
	\centering
	\includegraphics[clip,width=8.5cm]{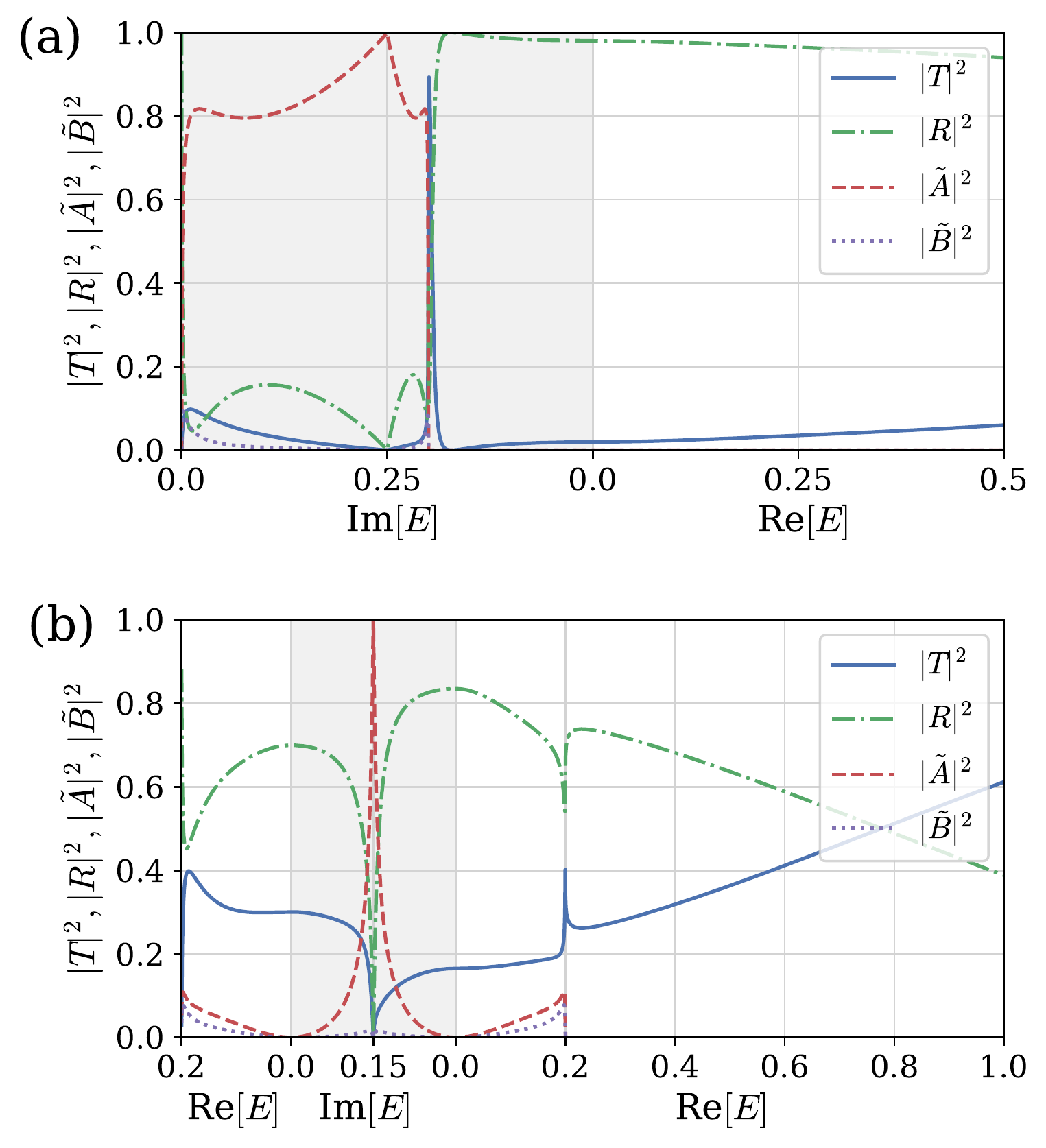}
			\caption{Tunneling property of the spin-wave modes in a polar BEC for (a) $E$ at $\beta=-0.25$, $q_z=0.1$, $U_0=3$ and $\sigma=0.5$, and (b) $\beta=-0.15$, $q_z=-0.1$, $U_0=2$ and $\sigma=0.5$.
			The horizontal axis is taken in the same manner as in Fig.~\ref{fig:binary_coeff}(b). The region where $E$ is pure imaginary is shaded with gray.
			The transmission probability $|T|^2$ is always smaller than unity, and its actual value depends on the detail of the barrier potential.
			At $E=i|\beta|$ in both panels, the perfect reflection ($|\tilde{A}|^2=1$) occurs.
			The transmission probability resonantly increases at $E=\sqrt{q_z(q_z+2\beta)}$, which corresponds to $E=0.2i$ in (a) and $E=0.2$ in (b).
			}
	\label{fig:polar_coef}
\end{figure}

In the presence of the quadratic Zeeman effect, the energy dispersion is categorized into four types
as shown in Fig.~\ref{fig:Energy2_ep_polar}: (a)$q_z>0$ and $q_z+2\beta>0$, (b)$q_z(q_z+2\beta)\le 0$ and $q_z+\beta>0$, (c)$q_z(q_z+2\beta)\le 0$ and $q_z+\beta\ge 0$, and (d)$q_z<0$ and $q_z+2\beta<0$.
We calculate the reflection and transmission probabilities for each case with changing the energy and momentum of the incident wave and find that the perfect transmission does not occur except for $q_z=0$.
Figures~\ref{fig:polar_coef}(a) and \ref{fig:polar_coef}(b) shows the behavior of the coefficients for the cases of Fig.~\ref{fig:Energy2_ep_polar}(c) and \ref{fig:Energy2_ep_polar}(d), respectively, where we set the horizontal axis in the same manner as Fig.~\ref{fig:binary_coeff}(b), i.e., $\epsilon_{k_{\rm in}}$ is zero at the left end of the figure and $\epsilon_{k_{\rm in}}$ increases as one goes right.
We find that $|T|^2<1$ for all region of the figure.
This result is consistent with the fact that the perfect transmission occurs when the quasiparticle wave function coincides with the condensate wave function: Since the quadratic Zeeman effect breaks the spin rotational symmetry, the spin-wave mode is not the NG mode of the system.
On the other hand, one can see that the perfect reflection occurs at the maximum $|{\rm Im} E|$ in both figures [at $E=0.25i$ and $E=0.15i$ for Figs.~\ref{fig:polar_coef}(a) and \ref{fig:polar_coef}(b),respectively], indicating that this is a universal property of dynamically unstable modes.

We note that the transmission probability resonantly increases at $E=0.2i$ in Fig.~\ref{fig:polar_coef}(a) and at $E=0.2$ in Fig.~\ref{fig:polar_coef}(b). These points correspond to the incident energy $E=\sqrt{q_z(q_z+2\beta)}$ with nonzero $k_{\rm in}$, which are depicted in Figs.~\ref{fig:Energy2_ep_polar}(c) and \ref{fig:Energy2_ep_polar}(d) with red circles. At these points, the momentum of the $A$ and $B$ terms in the asymptotic form becomes zero. Namely, the propagating modes for ${\rm Im}E>0.2$ [${\rm Re}E<0.2$] change into localized modes for ${\rm Im}E<0.2$ [${\rm Re}E>0.2$] in Fig.~\ref{fig:Energy2_ep_polar}(c) [\ref{fig:Energy2_ep_polar}(d)]. The increase in $|T|^2$ at this point is understood as a resonance with these $A$ and $B$ terms.
The peak value of the transmission probability depends on barrier potential and gets lower with increasing barrier potential.

\section{Conclusion\label{sec:conclusion}}
We have studied the tunneling properties of spin-wave mode in a dynamically unstable BEC and numerically shown that the perfect transmission occurs even in the presence of dynamical instability if the spin-wave is the NG mode.
When the prepared BEC is dynamically unstable, the eigenvalue $E$ of the BdG equation can be complex.
In the models we discussed, $E$ becomes pure imaginary for the incident momentum in a specific region and becomes zero when it changes from real to pure imaginary.
We have found that the perfect transmission occurs in the limit that both the eigenvalue $E$ and the momentum of the injected quasiparticle goes to zero.
This is the condition for the quasiparticle
to have the same form as the condensate wave function.
When the above condition is satisfied,
even a dynamically unstable mode that has a pure-imaginary $E$ exhibits the perfect transmission in the limit of $E\to 0$.
On the other hand, even when $E=0$, the perfect transmission does not occur when the incident momentum is nonzero.

Apart from the perfect transmission at $E\to 0$, 
we have also unveiled that the perfect reflection occurs at a point where $|{\rm Im}E|$ becomes its maximum.
Around the maximum of $|{\rm Im}E|$, there is a reflected wave that has a different momentum from the incident one.
At the occurrence of the perfect reflection, the incident wave destructively interferes with this reflected wave, and the quasiparticle wave function is strongly suppressed.
This result suggests that the most unstable mode that has the largest ${\rm Im} E$ cannot grow in the vicinity of a barrier potential.
The details of such instability dynamics remain as a future issue.
We have also found that the transmission probability resonantly increases when the reflected wave changes to a bound state around the potential barrier.

\begin{acknowledgment}

\vspace{2mm}
\noindent
{\bf Acknowledgment}
\vspace{1mm}

The authors thank Kazuya Fujimoto, Shun Tamura, and Ryoi Ohashi for fruitful discussions.
This work was supported by JST-CREST (Grant No. JPMJCR16F2) and JSPS KAKENHI
(Grants No. JP18K03538 and No. JP19H01824).

\end{acknowledgment}

\appendix
\section{Normalization condition for the Bosonic BdG equation}
\label{sec:append_normBdG}

The left and right eigenvectors of a non-Hermitian matrix $H$ are defined by
\begin{subequations}
\begin{align}
    H|w_n^{\rm R}\rangle=E_n|w_n^{\rm R}\rangle,
    \label{eq:eigenvector_R}\\
    \langle w_n^{\rm L}|H=\langle w_n^{\rm L}|E_n.
\end{align}
The second equation can be rewritten as
\begin{align}
    H^\dagger|w_n^{\rm L}\rangle=E_n^*|w_n^{\rm L}\rangle,
    \label{eq:eigenvector_L}
\end{align}
\end{subequations}
where $|w_n^{\rm L}\rangle$ is the Hermite conjugate of $\langle w_n^{\rm L}|$.
When we multiply $\langle w_m^{\rm L}|$ to Eq.~\eqref{eq:eigenvector_R} form the left, we obtain
\begin{align}
    \langle w_m^{\rm L}|H|w_n^{\rm R}\rangle = \langle w_m^{\rm L}|E_n|w_n^{\rm R}\rangle &= \langle w_m^{\rm L}|E_m|w_n^{\rm R}\rangle,\\
    (E_m-E_n)\langle w_m^{\rm L}|w_n^{\rm R}\rangle &= 0,
\label{eq:orthnorm}
\end{align}
from which the orthonormal condition is given by
\begin{align}
    \langle w_m^{\rm L}|w_n^{\rm R}\rangle=\delta_{mn}.
\end{align}

For the case of a bosonic BdG equation, the matrix $H$ satisfies
the pseudo-Hermiticity and the particle-hole symmetry:
\begin{align}
    \tau_z H \tau_z &= H^\dagger,
    \label{eq:pseudo-Hermiticity}\\
    \mathcal{C}^{-1} H \mathcal{C} &= -H,
    \label{eq:PHS}
\end{align}
where $\mathcal{C}\equiv \tau_x K$ is the particle-hole operator with $K$ being the complex-conjugate operator and $\tau_{x,y,z}$ the Pauli matrices in the Nambu space.
From Eqs.~\eqref{eq:eigenvector_L} and \eqref{eq:pseudo-Hermiticity},
we obtain
\begin{align}
    H\tau_z|w_n^{\rm L}\rangle = E_n^* \tau_z|w_n^{\rm L}\rangle,
\label{eq:pH_eigenvector}
\end{align}
which leads to $|w_n^{\rm L}\rangle \propto \tau_z|w_n^{\rm R}\rangle$ for real $E_n$.
From Eqs.~\eqref{eq:eigenvector_R} and \eqref{eq:PHS}, we obtain
\begin{align}
    H\mathcal{C}|w_n^{\rm R}\rangle = -\mathcal{C}H|w_n^{\rm R}\rangle = -E_n^*\mathcal{C}|w_n^{\rm R}\rangle.
    \label{eq:PHS_eigenvector}
\end{align}

When $E_n$ is real, $|w_n^{\rm R}\rangle$ and $\mathcal{C}|w_n^{\rm R}\rangle$ are a particle-hole pair and satisfy
\begin{align}
\langle w_n^{\rm R}|\mathcal{C}\tau_z \mathcal{C} |w_n^{\rm R}\rangle=
-\langle w_n^{\rm R}|\tau_z|w_n^{\rm R}\rangle,
\end{align}
which implies $|w_n^{\rm L}\rangle=\tau_z|w_n^{\rm R}\rangle$ ($|w_n^{\rm L}\rangle=-\tau_z|w_n^{\rm R}\rangle$) for a particle (hole) mode.
We therefore define the normalization constant for a real-eigenvalue mode as
\begin{align}
    \langle w_n^{\rm R}|\tau_z|w_n^{\rm R}\rangle=1\ \textrm{or}\ -1.
\label{eq:norm_r_general}
\end{align}

For the case when ${\rm Im}\,E_n\neq 0$,
there exists $n'$ such that $E_{n'}=E_n^*$, $|w_{n'}^{\rm R}\rangle \propto \tau_z |w_n^{\rm L}\rangle$, and  $|w_{n}^{\rm R}\rangle \propto \tau_z |w_{n'}^{\rm L}\rangle$ [see Eq.~\eqref{eq:pH_eigenvector}].
Then, the normalization condition is given by $|\langle w_{n'}^{\rm R}|\tau_z|w_n^{\rm R}\rangle|=1$.
As a special case, 
when the matrix elements of $H$ are all real, we obtain $|w_{n'}^{\rm R}\rangle=\left(|w_n^{\rm R}\rangle\right)^*$,
from which the normalization condition is given by
\begin{align}
    \left|\left( \langle  w_n^{\rm R}|\right)^*\tau_z|w_n^{\rm R}\rangle\right|=1.
    \label{eq:norm_c_general}
\end{align}

The BdG equations discussed in this paper have only real ($\in\mathbb{R}$) or pure imaginary ($\in i\mathbb{R}$) eigenvalues.
When the BdG equation is written in the Fourier space as in Eqs.~\eqref{eq:BdG_binary_magnon_k}, the normalization condition is given by
\begin{subequations}
\begin{align}
    |u|^2-|v|^2&=\pm 1 \ \ (E\in \mathbb{R}),\\
    u^2-v^2&=1 \ \ \ \ \,(E\in i\mathbb{R}),
\end{align}
\end{subequations}
where the second equation also determines the phase of the eigenvector.
When the BdG equation is written in the coordinate space as in Eqs.~\eqref{eq:BdG_binary_magnon_x} and \eqref{eq:BdG_spin1_magnon_x}, Eq.~\eqref{eq:orthnorm} means that  
\begin{align}
    \displaystyle{\int_{-\infty}^\infty \left[|u(x)|^2-|v(x)|^2\right]dx}
\end{align}
can be regarded as a norm for $E\in\mathbb{R}$
whereas it always vanishes for $E\in\mathbb{C}$.

\bibliography{reference.bib}
\bibliographystyle{jpsj}

\end{document}